\documentclass[twocolumn,american,aps,prb,showpacs,showkeys,amsmath,amssymb,superscriptaddress,a4]{revtex4-1}
\usepackage[T1]{fontenc}
\usepackage[latin9]{inputenc}
\usepackage{array}
\usepackage{float}
\usepackage{multirow}
\usepackage{amsmath}
\usepackage{amssymb}
\usepackage{graphicx}
\usepackage{subscript}

\usepackage{color}

\newcommand {\dd} {\text{d}}

\makeatletter

\providecommand{\tabularnewline}{\\}

\usepackage{multirow}\usepackage{bm}\usepackage{gensymb}\usepackage{threeparttable}

\makeatother

\usepackage{babel}
\begin{document}
\selectlanguage{american}%

\title{A fast anharmonic free energy method with an application to vacancies
in ZrC}

\author{Thomas A. Mellan}
\email{t.mellan@imperial.ac.uk}
\affiliation{Thomas Young Centre, Department of Physics and Department of Materials, Imperial College London, Exhibition Road, London SW7
2AZ, United Kingdom}

\author{Andrew I. Duff}

\affiliation{STFC Hartree Centre, Scitech Daresbury, Warrington WA4 4AD, United
Kingdom}

\author{Blazej Grabowski }

\affiliation{Max-Planck-Institut f{\"u}r Eisenforschung GmbH, D-40237 D{\"u}sseldorf,
Germany}

\author{Michael W. Finnis}

\affiliation{Thomas Young Centre, Department
of Physics and Department of Materials, Imperial College London, Exhibition
Road, London SW7 2AZ, United Kingdom}

\date{\today}

\selectlanguage{american}%
\begin{abstract}
We propose an approach to calculate the anharmonic part of the volumetric-strain
and temperature dependent free energy of a crystal. The method strikes
an effective balance between accuracy and computational efficiency,
showing a $\times10$ speed-up on comparable free energy approaches at the level of density functional theory,
with average errors less than 1 meV/atom. As a demonstration
we make new predictions on the thermodynamics of substoichiometric
ZrC$_x$, including vacancy concentration and heat capacity.
\end{abstract}

\maketitle

\section{Introduction \label{sec:Introduction}}

Thanks to recent advances in computational thermodynamics, the thermal
properties of metals such as aluminium and gold have been investigated
up to the melting point, using thermodynamic integration (TI) with Langevin dynamics\cite{Grabowski2009,Grabowski2015,Glensk2014} based on density functional theory (DFT).
A two-step TI approach increases computational efficiency further,
making predictions possible for more complex materials at the DFT level
of theory. Examples so far include the ultra-high-temperature ceramics
ZrC ($T_{m}=3700\,\text{K}$) and HfC ($T_{m}=4160\,\text{K}$),\cite{Duff2015,Duff2018} and recent attempts at tackling the emerging class of multicomponent systems.\cite{Ikeda2018,Grabowski} 
Such calculations are not yet routine, but the course of our research and the recent methodological developments of others\citep{Wu2009,Moustafa2017a,Purohit2018} in this field is 
in that direction. Here we present some new developments that are a step towards the goal of routinely               
computing accurate free energies for hard matter systems, including
binaries, ternaries, and high-entropy alloys, across  the 
range of temperatures, pressures and chemical potentials, up to and eventually beyond the melting point.

In this work we compute the concentration of vacancies in ZrC$_x$
and associated ambient pressure thermodynamics for small deviations from stoichiometry. The
ZrC$_x$ free energy and derivatives are analyzed in
terms of the contributions\citep{Zhang2018}
\begin{equation}
F=E_{0}+F_{\text{el}}+F_{\text{qh}}+F_{\text{ah}}+F_{\text{el-vib}}+F_{\text{config}},\,\label{eq: free_energy_contributions}
\end{equation}
in which $E_{0}$ is the  DFT energy of a static lattice at $T=0\,$K, $F_{\text{el}}$ is the Helmholtz free energy contribution from the thermal excitations of electrons, $F_{\text{qh}}$ is the quasiharmonic vibrational contribution, $F_{\text{ah}}$ is the anharmonic vibrational contribution, $F_{\text{el-vib}}$ is the electron-vibration contribution, and $F_{\text{config}}$ is the contribution of configurational entropy due to the number of distinct point-defect distributions.
For each of the five temperature-dependent terms we have calculated the dependence on the independent variables volume and temperature up to the melting point. Considerable attention in this paper is given to the method we use
to compute the challenging anharmonic term, $F_{\text{ah}}$. The
method described achieves effective DFT accuracy in $F_{\text{ah}}$ (1 meV
per bulk atom) with only an order of magnitude greater computational
cost than ordinary quasiharmonic free energy calculations. 

This paper is set out as follows. Sec.\,\ref{subsec: methodBackground} gives the context of
our approach, theoretical
details are in Sec.\,\ref{subsec: MTILDapproach}, a description of the modified embedded atom method (MEAM) potential fitting, which reduces overall the number of expensive DFT calculations, 
in Sec.\,\ref{subsec:Potential-fitting}, thermodynamic integration
in Sec.\,\ref{subsec:Expectations-and-uncertainties}, and DFT technical
details in Sec.\,\ref{subsec:Technical}. Benchmarking is described
in terms of accuracy and precision in Sec.\,\ref{subsec: benchMarksAccuracy}
and computational efficiency in Sec.\,\ref{subsec: speedUp}. Application
to ZrC$_x$ provides insight into the nature of anharmonicity
in substoichiometric binary crystals in Sec.\,\ref{subsec: vacancyAnharmonicity},
prediction of vacancy concentration in Sec.\,\ref{subsec: anharmonicVacancyEnergy},
and analysis of ZrC$_x$ heat capacity in Sec.\,\ref{subsec: testApplicationZrCx-heatCapacity}.

\section{Methods\label{sec:Method}}

\subsection{Background\label{subsec: methodBackground}}

There are a number of approaches in the literature to calculate the  anharmonic vibrational properties of crystals,\cite{Alfe2001,Alfe2002a,Alfe2002b,Ackland2002,Duff2015,Errea2011,Glensk2014,Grabowski2009,Grabowski2015,Hellman2013,Monserrat2013a,Klein1972a,Fn1964,Frenkel1984,Moustafa2017}
including thermodynamic integration\cite{Kirkwood1935} (TI), which is the method used in this work.  In TI 
the anharmonic part of the full Hamiltonian, $E-E_{\text{qh}}$, is switched on with the parameter
$\lambda\in[0,\,1]$, in this instance linearly as $E_{\text{mix}}(\lambda)=E_{\text{qh}}+\lambda(E^{\text{}}-E_{\text{qh}}^{\text{}})$.
Classical averages of $\partial_{\lambda}E_{\text{mix}}(\lambda)$
are obtained stochastically from molecular dynamics (MD), and numerically integrated along
the coupling path to give the free energy due to $E_{\text{ah}}$:

\begin{equation}
F_{\text{ah}}=\int_{0}^{1}d\lambda\,\left\langle \partial_{\lambda}E_{\text{mix}}(\lambda)\right\rangle _{\lambda}\,.\label{eq: basicAnharmonicFreeEnergy}
\end{equation}
Note, $\partial_{\lambda}E_{\text{mix}}(\lambda)=E(\mathbf{R},\,V)-E_{\text{qh}}(\mathbf{R},\,V)$,
where $E(\mathbf{R},\,V)$ is the full potential energy surface
and $E_{\text{qh}}(\mathbf{R},\,V)$ the volume-dependent harmonic potential energy surface in Born-Oppenheimer nuclear coordinates $\mathbf{R}$.

The partitioning of the vibrational free energy $F_{\text{vib}}=F_{\text{qh}}+F_{\text{ah}}$ divides
the problem conveniently into a simple quantum mechanical part, in which the vibrations are quantised as phonons, and an anharmonic part in which the vibrations are treated classically.  Thus $F_{\text{vib}}$  has the appropriate low-temperature quantum
statistics. The anharmonicity is treated
classically but also non-perturbatively, which is important at high
temperatures as the melting point is approached. In order to evaluate the anharmonic term,
the expectation values $\left\langle \partial_{\lambda}E_{\text{mix}}(\lambda)\right\rangle _{\lambda}$
require  between $10^3 \ldots 10^7$ configurations for a typical $\lambda$-ensemble
at a typical supercell size. To produce a free energy surface $F_{\text{ah}}(V,\,T)$
one must sample ensembles across dimensions of strain (here volume),
temperature and coupling parameter, $N_{\lambda}\times N_{V}\times N_{T}\approx10^2 \ldots 10^3$.
Thus the ball-park number of total energy calculations, between $10^5 \ldots 10^{10}$
configurations, is prohibitive at the highly-converged DFT level of
accuracy required.

In one approach to reduce computational complexity, $F_{\text{ah}}$
is obtained by cumulating a sequence of thermodynamic integrations. In an implementation
of this approach referred to as TU-TILD,\cite{Duff2015} which is expressed by Eqn.~(\ref{eq: TU-TILD}), much of $F_{\text{ah}}$
is captured using an inexpensive MEAM potential. This results in faster convergence of the expensive TI from
MEAM to DFT, expressed in the last term of Eqn. (\ref{eq: TU-TILD}).
\begin{align}
F_{\text{ah}}^{\text{TUTILD}} & =\!\!\int_{0}^{1}\!\!\dd\lambda\!\left\langle E^{\text{DFT}}(\mathbf{R},V)-E_{\text{qh}}^{\text{DFT}}(\mathbf{R},V)\right\rangle _{\lambda}\nonumber \\
 & =\!\!\int_{0}^{1}\!\dd\lambda\!\left\langle E^{\text{MEAM}}(\mathbf{R},V)-E_{\text{qh}}^{\text{DFT}}(\mathbf{R},V)\right\rangle _{\lambda}\notag\\
 &\,\,+\!\!\int_{0}^{1}\!\!\dd\lambda\!\left\langle E^{\text{DFT}}(\mathbf{R},V)-E^{\text{MEAM}}(\mathbf{R},V)\right\rangle _{\lambda}\,.
 \label{eq: TU-TILD}
\end{align}
In practice to save computation time, the DFT MD calculations in a TU-TILD procedure were usually performed with a low-converged expansion of the wavefunctions, using a reduced number of plane-waves, and fewer k-points than required for maximum accuracy.  The maximum accuracy was then obtained by up-sampling, as in the original UP-TILD method\cite{Grabowski2009}.  
The methodology we introduce below, inspired by these approaches, was devised in order to make significant further savings in computation time without sacrificing accuracy.

\subsection{MEAM thermodynamic integration approach\label{subsec: MTILDapproach}}

The approach we propose in this work can be summarized by

\begin{equation}
F_{\text{ah}}^{\text{}}=\int_{0}^{1}d\lambda\,\left\langle E^{\text{MEAM}}(\mathbf{R},\,V)-E_{\text{qh}}^{\text{DFT}}(\mathbf{R},\,V)\right\rangle _{\lambda}\,.
\label{eq: fahEqn}
\end{equation}
Our approach calculates the anharmonic free energy of a MEAM crystal
referenced to a harmonic DFT crystal, which is formally the first stage in 
Eqn.~(\ref{eq: TU-TILD}). In the present method the quasiharmonic Helmholtz free energy at each volume is still explicitly represented by the volume-dependent dynamical matrix calculated with DFT, which captures much of the thermal expansion, but the anharmonic terms are now entirely described by the MEAM potential. The success of the method depends on being able to generate a MEAM potential of sufficient accuracy to replace the anharmonic DFT contribution.
It is by no means obvious \emph{a priori} that this is possible, or if it is, that the process of generating the potential is not too expensive to 
warrant the effort.

From the potential terms in Eqn.(\ref{eq: fahEqn}) it is clear that $F_{\text{ah}}^{\text{}}$ can be evaluated by this method to a high level of precision using modest
computational resources, but the MEAM TI approximation introduces
systematic potential errors with respect to DFT TI. Accuracy must
be carefully controlled by generating custom MEAM potentials from
high-quality DFT MD. Generating 
the training and validation data is time consuming,  so fitting the potentials becomes
the  primary computational cost in predicting $F_{\text{ah}}$
in our TI approach. These costs incurred before doing the
MEAM TI will be shown in Sec. \ref{subsec: benchMarksAccuracy} to
be comfortably small enough. Details of potential fitting
and error control are presented in the following section.

\subsection{Potential fitting\label{subsec:Potential-fitting}}
\begin{center}
\begin{figure*}
\begin{centering}
\includegraphics[scale=0.55]{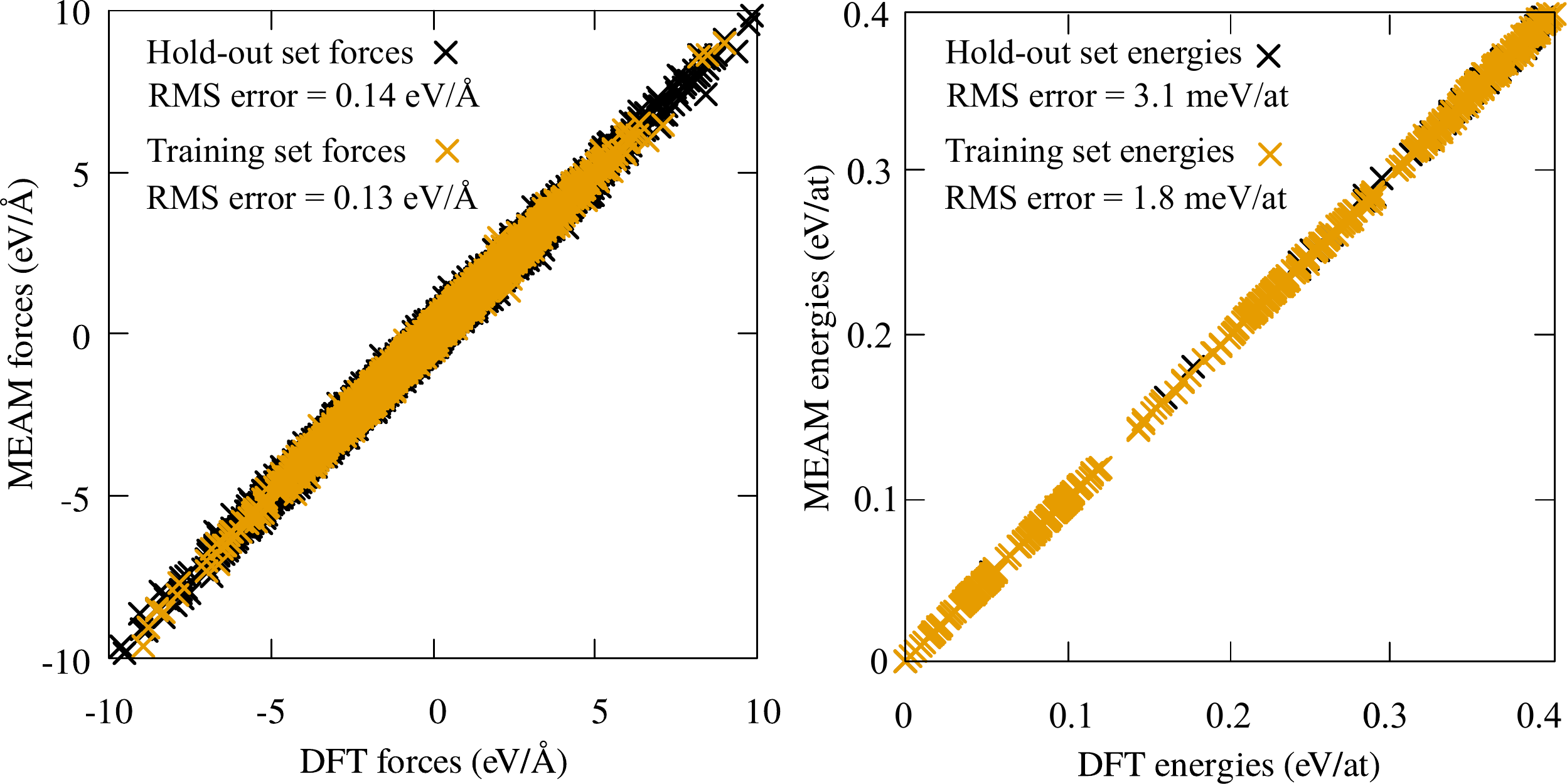}
\par\end{centering}
\caption{\foreignlanguage{american}{Quality of interatomic potential: MEAM versus DFT forces and energies
shown for training and hold-out Zr\protect\textsubscript{32}C\protect\textsubscript{31}
data sets. \label{fig: pltMEAMfitting}}}
\end{figure*}
\par\end{center}

Interatomic potentials have been fitted using the reference-free modified
embedded atom method (RF-MEAM).\cite{Duff2015a} 
The MEAM potentials fitted in this work lack transferability and are specialized to perform for the intended application. For instance, to model $F_{\text{ah}}(V,\,T)$ we fit a separate
potential at each volume considered, which minimizes the possibility
of strain-dependent errors. Obtaining the correct implicit
volume dependence of a potential is important, as explicit anharmonic effects depend sensitively on the degree of lattice expansion, as demonstrated in Sec. \ref{subsec: vacancyAnharmonicity}. 

Potentials for Zr\textsubscript{32}C\textsubscript{32} and Zr\textsubscript{32}C\textsubscript{31}
have been fitted at the cell lattice parameters  $a=\{4.685,\,4.730,\,4.759,\,4.801,\,4.850\}$
\AA. In crystals of lower than cubic symmetry, thermal expansion may involve other modes of strain, but in our case only volumetric strain need be considered. At each volume the potential is fitted to configurations sampled
from DFT MD runs between $T=200$\,K and $T=3800$\,K. The fitting  set for
each volume comprises $10^{3}$ Zr\textsubscript{32}C\textsubscript{31}
configurations, which supply energies and forces to the objective
function, for minimizing force residuals and energy-residual variances.
Fitting to the forces on  each atom allows the potential to be specified by fewer distinct
configurations than fitting to energies, as $N_{\text{forces}}=3N_{\text{atom}}N_{\text{energies}}$.
To generate the fitting data points efficiently, low-quality DFT can
be up-sampled to high-quality DFT to produce high-quality target
forces and total energies.

The interatomic potentials are generated using a genetic-algorithm conjugate-gradient
fitting procedure implemented and publicly available in the \textsc{\footnotesize{}MEAMFIT2}
code.\cite{Duff2015a,Duff2016,MEAMfitStatement} The code fits an
RF-MEAM potential that permits locally positive and negative density
terms in order to increase variational freedom, subject to a net positive background.
The fitted potential has 3 embedding and 3 pairwise terms, within
a radial cutoff of $4.8$\,\AA, which includes interactions up to third nearest-neighbor.
These potential parameters provide a satisfactory compromise between
accuracy and complexity, in terms of minimizing residual variances
on hold-out data using the fewest degrees of freedom (78 parameters
for the 3-3 potential). The quality of fits for energy and forces is presented
in Fig.~\ref{fig: pltMEAMfitting} for Zr\textsubscript{32}C\textsubscript{31}. 

In this RF-MEAM application a different potential is fitted at each
volume but we require a potential to be transferable  across composition, i.e. we want
the same potential to describe both Zr\textsubscript{32}C\textsubscript{31}
and Zr\textsubscript{32}C\textsubscript{32}, for a given volume at any temperature.
This  transferability ensures a systematic 
error cancellation in $F_{\text{ah}}$ for Zr\textsubscript{32}C\textsubscript{31}
and Zr\textsubscript{32}C\textsubscript{32} that helps in calculating accurate fully anharmonic vacancy formation energies. 
\subsection{Thermodynamic integration \label{subsec:Expectations-and-uncertainties}}
\begin{center}
\begin{figure}[H]
\begin{centering}
\includegraphics[scale=0.55]{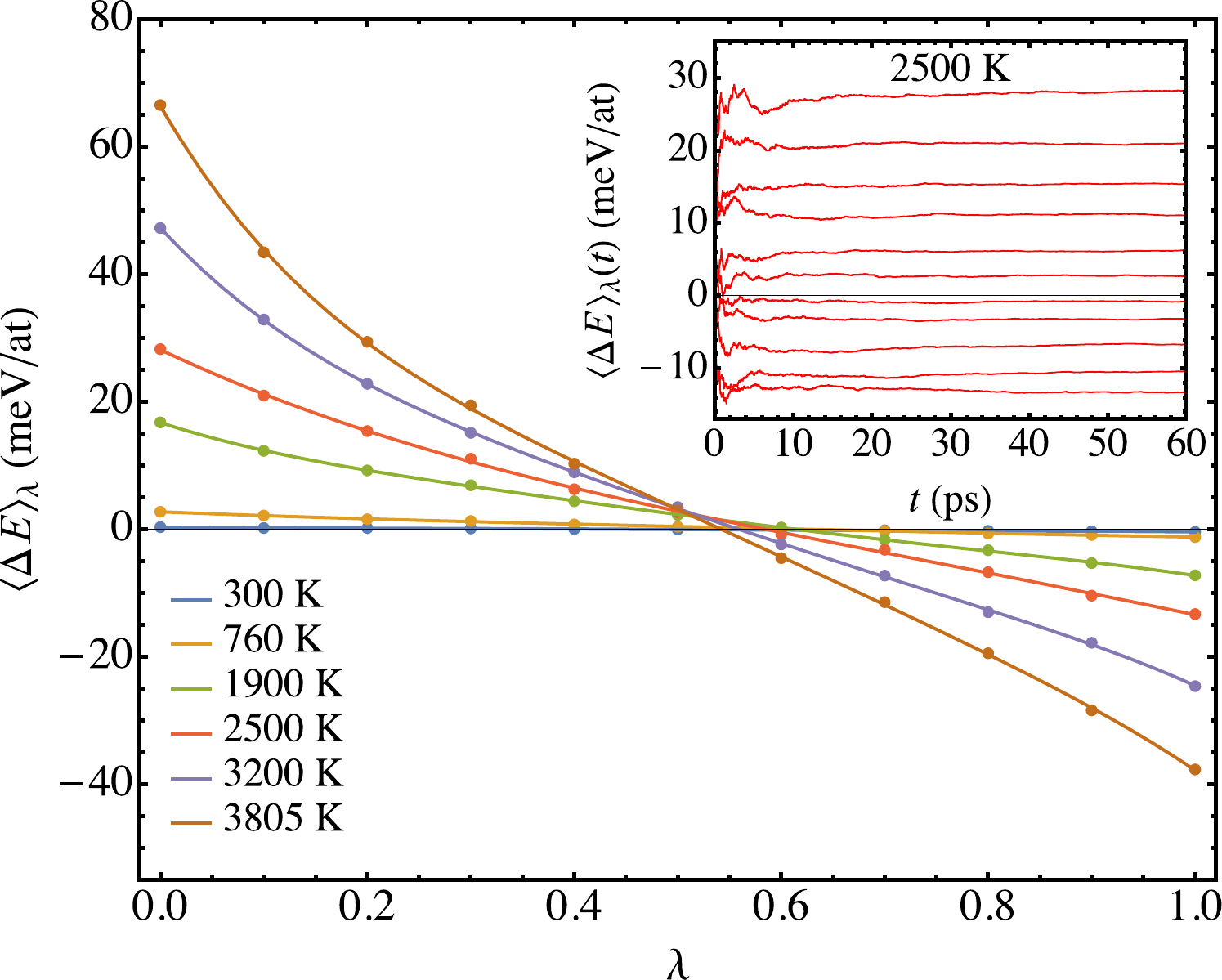}
\par\end{centering}
\caption{\foreignlanguage{american}{Thermodynamic integrand $\left\langle \Delta E\right\rangle _{\lambda}(\lambda)$
for Zr\protect\textsubscript{32}C\protect\textsubscript{31} at $a=4.801$
\AA. \emph{Inset}: Convergence of $\left\langle \Delta E\right\rangle _{\lambda}(t)$
with MD time-step, at ensembles $ \lambda_{i} =i/10$.
\label{fig: pltdUdL}}}
\end{figure}
\par\end{center}

$F_{\text{ah}}$ is estimated by computing $\left\langle \Delta E\right\rangle _{\lambda}$, with $\Delta E= E^\text{MEAM} - E^\text{DFT}_\text{qh}$,
for a series of $11$ ensembles at equal increments of $\lambda$,  $ \lambda_{i}=i/10$.
In Fig.~\ref{fig: pltdUdL} we show the dependence of $\left\langle \Delta E\right\rangle _{\lambda}$
on $\lambda$ across a series of temperatures.
We see that at each temperature, $\partial_{\lambda}\langle\Delta E\rangle_\lambda \le 0$, a necessary condition that  is easy to prove, as in a derivation of the Gibbs-Bogoliubov inequality.
In Fig. \ref{fig: pltdUdL} inset the convergence
of $\left\langle \Delta E\right\rangle _{\lambda}$ is shown for the
first 60,000 time steps of a simulation. Expectation values are generated
using Langevin MD, with a one femtosecond time-step,\cite{Duff2015}
and a friction parameter of $\gamma=0.05$\,fs\textsuperscript{-1}
for Zr\textsubscript{32}C\textsubscript{31} and $\gamma=0.01$\,fs\textsuperscript{-1}
for Zr\textsubscript{32}C\textsubscript{32}\@. At each pair, $\{V_{i}\,T_{i}\}$,
the expectation value $\left\langle \Delta E\right\rangle _{\lambda}$
is fitted in $\lambda$ by least squares to the truncated power series
\begin{equation}
\left\langle \Delta E\right\rangle _{\lambda}(\lambda)=\sum_{i=0}^{i=5}a_{i}\lambda^{i}\,,\,\,\lambda\in[0,\,1]\,,
\label{eq: deltaUexpectation}
\end{equation}
for which the coefficients are alternating in sign and converging. If intrinsic defects form and migrate on the atomic vibration time-scale this series is expected to poorly converge. In this work we exclude any system configurations in which Frenkel defects have spontaneously formed, for example at the melting point, in order to ensure well-converged thermodynamic integrations of the anharmonic free energy of defect-free ZrC$_x$.

\begin{center}
\begin{figure}[H]
\begin{centering}
\includegraphics[scale=0.60]{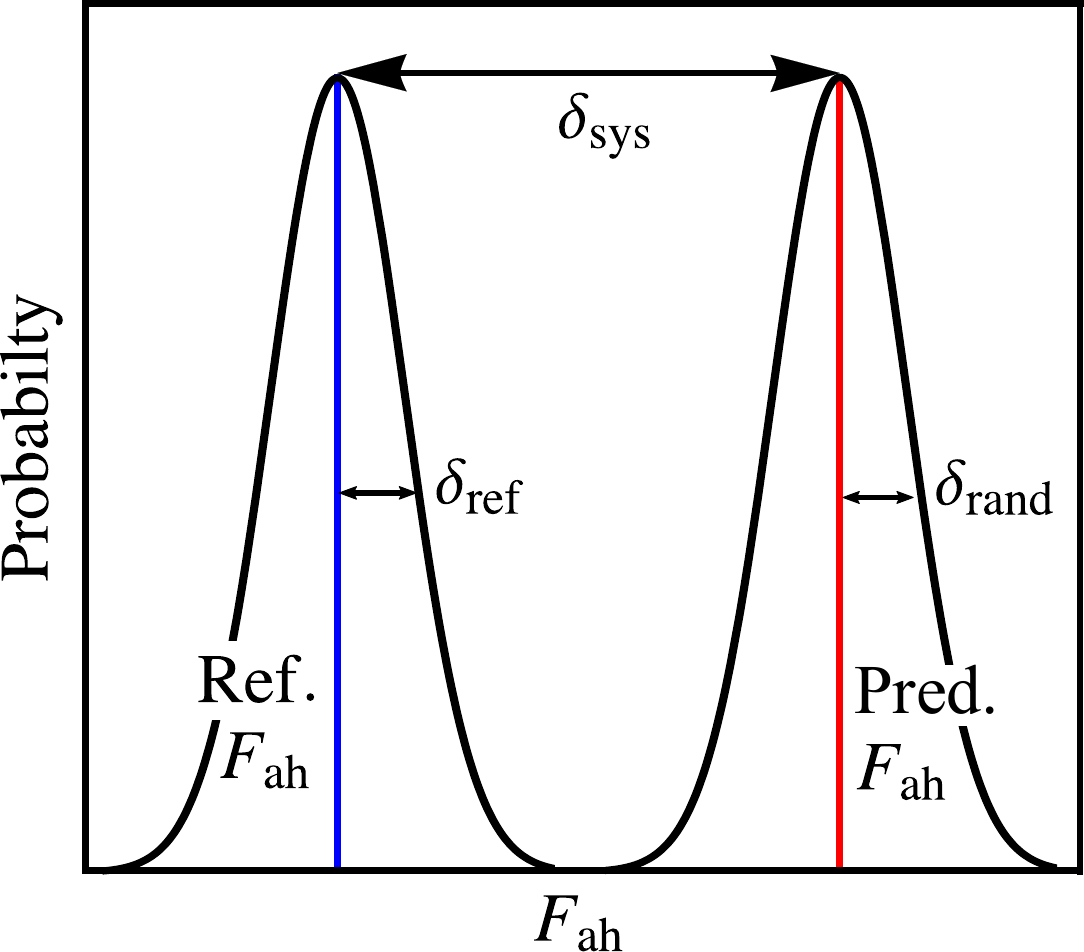}
\par\end{centering}
\caption{\foreignlanguage{american}{Schematic relating precision ($\delta_{\text{rand}}$) and accuracy
($\delta_{\text{sys}}$) errors in our predicted $F_{\text{ah}}$ value
to a reference $F_{\text{ah}}$ value. The benchmark value is from
TU-TILD\cite{Duff2015} and is converged to $\delta_{\text{ref}}=0.1$
meV/at. Precision $\delta_{\text{rand}}$ can easily be reduced to
0.1 meV/at or less, minimizing the systematic potential error. $\delta_{\text{sys}}$
is the main challenge. \label{fig: accuracy-preicision}}}
\end{figure}
\par\end{center}

Errors in  predicting $F_{\text{ah}}$ are considered from two
primary sources, namely statistical convergence and a systematic potential
error. The DFT benchmark also has a small convergence error which
is accounted for, but other sources of error, such as from DFT exchange-correlation, electron-phonon scattering, and other quantum effects beyond
the harmonic approximation, are beyond the scope of this paper. The
three countable contributions are shown schematically in Fig. \ref{fig: accuracy-preicision},
and give the total expected error of
\begin{equation}
\delta=\sqrt{\delta_{\text{sys}}^{2}+\delta_{\text{rand}}^{2}+\delta_{\text{ref}}^{2}}\,.
\label{eq: errorQuadrature}
\end{equation}
Precision error $\delta_{\text{rand}}$ arises from evaluating an
observable from a finite number of samples in the MEAM MD, and the
systematic error $\delta_{\text{sys}}$ is due to the energy difference
between a MEAM potential and DFT. The convergence error in the benchmark
$F_{\text{ah}}$ value from TU-TILD is $\delta_{\text{ref}}=0.1$
meV/at.\cite{Duff2015}

The statistical convergence $\delta_{\text{rand}}$ is computed using stratified systematic sampling.\cite{Haile1997} In the simulation the precision error scales as\cite{Janke2002}
\begin{equation}
\delta_{\text{rand}}\sim\sqrt{\frac{\sigma^2_{\lambda}}{N_\lambda} \frac{\tau}{t}}\,,
\label{eq: statisticalPrecisionError}
\end{equation}
where $N_\lambda$ is the number of integration path points sampled (with $N_\lambda=11$ in our case), $\sigma_{\lambda}$ is the norm of $\Delta E$ standard deviations,  $\tau$ is the $\Delta E$ autocorrelation time (ca. $11$ fs, see Appendix), and $t$ is the simulation time (ca. $0.1$ ns). We emphasise that the TI method described differs from other approaches in that statistical convergence is
not accuracy-limiting, for example, nanosecond simulations can comfortably be performed in a day on a low-performance computing platform.

The systematic potential error in the anharmonic free energy $\delta_{\text{sys}}$
can be computed by thermodynamic integration
\begin{equation}
\delta_{\text{sys}}=\int_{0}^{1}d\lambda\,\left\langle E^{\text{DFT}}(\mathbf{R},\,V)-E^{\text{MEAM}}(\mathbf{R},\,V)\right\rangle _{\lambda}\,.
\label{eq: anharmonicFreeEnergyTI}
\end{equation}
$\delta_{\text{sys}}$ is the primary error source in the method we
describe to compute $F_{\text{ah}}$. Accuracy benchmarks in Sec.
\ref{subsec: benchMarksAccuracy} show $\delta_{\text{sys}}$ can
be sufficiently controlled to satisfy 1 meV/at bulk convergence across
the $F_{\text{ah}}(V,\,T)$ surface. 

\subsection{Technical details\label{subsec:Technical}}

Periodic plane-wave DFT calculations were
performed using the VASP software,\cite{Kresse1996,Kresse1996a} with
the local density approximation (LDA) exchange-correlation function.\cite{Perdew1981a}
The projector-augmented wave (PAW) method was used,\cite{Kresse1999}
with $4s$- and $4p$-Zr electrons included as valence states.

$E_{0}$ was computed on a mesh of 11 volumes, and at each volume the
internal coordinates have been relaxed to give residual forces under  $10^{-6}$ eV/\AA. 
Self-consistent field (SCF) 
total
energies and energy eigenvalues have been resolved to $10^{-9}$ eV. Methfessel-Paxton
smearing has been used with a width of 0.1 eV.\cite{Methfessel1989}
The kinetic energy cutoff has been set to 700 eV and k-point mesh
$12\times12\times12$ for the $2\times2\times2$ supercell. The $E_{0}$
vacancy formation energy contribution is extrapolated to the dilute
limit, using data points from Zr\textsubscript{32}C\textsubscript{31},
Zr\textsubscript{108}C\textsubscript{107} and Zr\textsubscript{256}C\textsubscript{255}.

For the quasiharmonic Helmholtz free energy $F_{\text{qh}}$, the kinetic energy cutoff has been set to 700
eV and k-point mesh to $6\times6\times6$ for the $2\times2\times2$
supercell. 
Phonons were calculated using the small displacement supercell
method with the \textsc{\footnotesize{}PHONOPY} code.\cite{Togo2015a}
At each of the 11 $2\times2\times2$ supercells that span the range of lattice parameters $[4.575,\,4.875]$ \AA, 
sets of 18 displacements
were made for Zr\textsubscript{32}C\textsubscript{31} and sets of
four displacements for Zr\textsubscript{32}C\textsubscript{32}\@.
The phonon q-points were sampled by a mesh of $25\times25\times25$
points for the $2\times2\times2$ supercells.

The electronic Helmholtz free energy $F_{\text{el}}(V,\,T)$ has been
calculated using the Mermin finite-temperature formulation of DFT,\cite{Mermin1965}
on a mesh of 10 temperatures and 8 volumes sampled between $V_{\text{eq}}(T=0\,\text{K})$
and $V_{\text{eq}}(T_{m})$. Electron states are self-consistent to
at least $10^{-7}$ eV/atom. We used $384$ bands, which was sufficient
to span all states with partial occupation up to the melting point $T_{m}$.
A kinetic cutoff energy of 700 eV was used, with k-point sampling
at $12\times12\times12$ for the $2\times2\times2$ supercells. 

The electron-vibration Helmholtz free energy $F_{\text{el-vib}}(V,\,T)$, has been calculated from low-converged MD configurations that are subsequently up-sampled, as in the proceedure recently performed for a number of transition metals.\cite{Zhang2017} The electronic free energy is calculated for each MD configuration, using the Mermin formulation at electronic temperature corresponding to the MD ensemble temperature. At each volume-temperature mesh point, the electronic free energies are averaged over the ensemble configurations, and referenced to the perfect crystal, in order to find the electron-vibration coupling contribution to the Helmholtz free energy.

The anharmonic Helmholtz free energy $F_{\text{ah}}(V,\,T)$ was determined
using a mesh of six temperatures and five volumes. Temperatures span
$0\,$K to $T_{m}$, and volumes $V_{\text{eq}}(T=0\,\text{K})$ to $1.15V_{\text{eq}}(T_{m})$.
Potentials were fitted to MD configurations from DFT that used a 700
eV cutoff and k-point sampling mesh of $6\times6\times6$ for the
$2\times2\times2$ supercell.

\clearpage{}

\section{Benchmarks \label{sec: BenchMarks} }

\subsection{Accuracy and precision\label{subsec: benchMarksAccuracy}}
\begin{center}
\begin{figure}[H]
\begin{centering}
\includegraphics[scale=0.4]{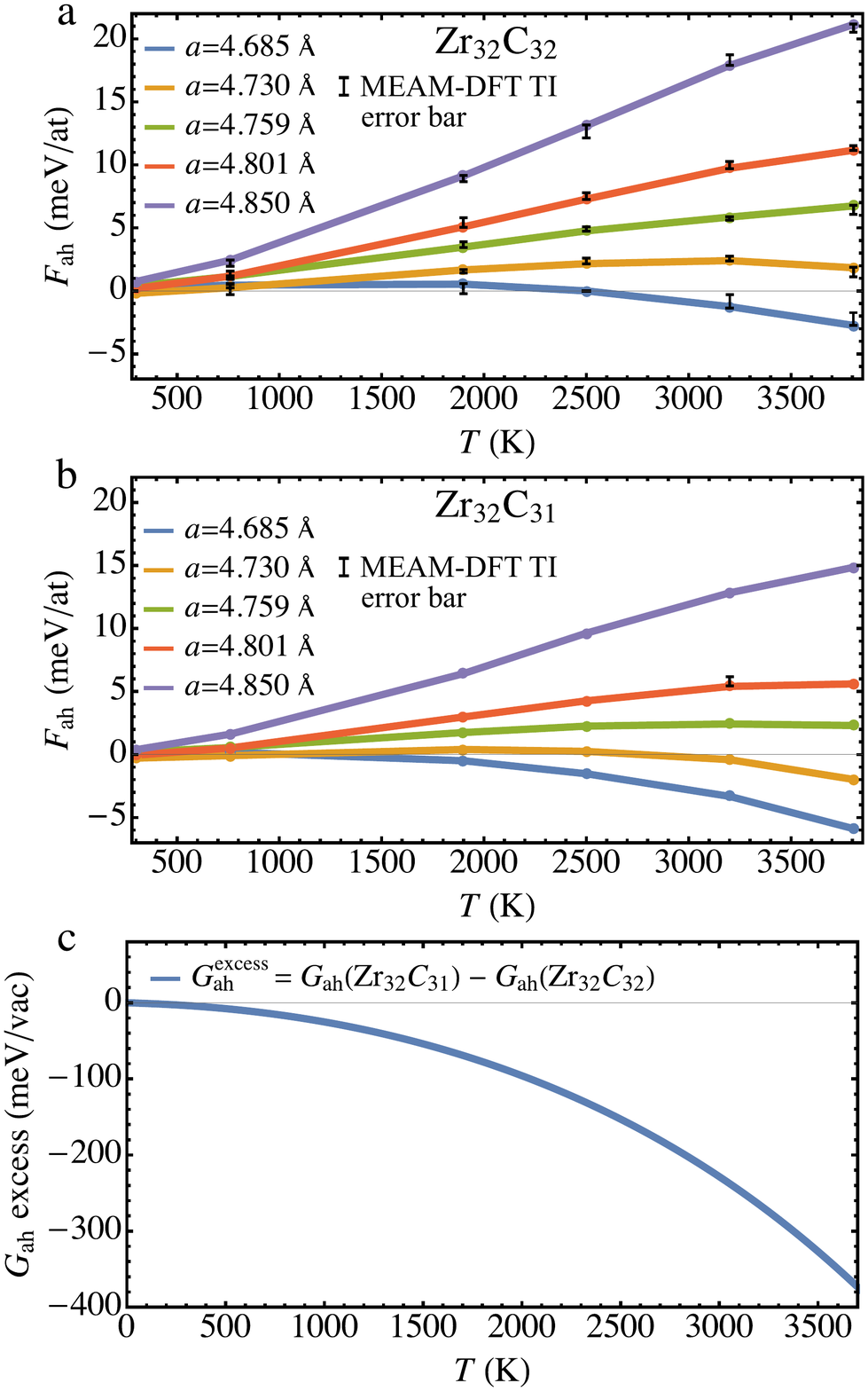}
\par\end{centering}
\caption{\foreignlanguage{american}{a) Anharmonic Helmholtz free energy $F_{\text{ah}}$ for perfect Zr\protect\textsubscript{32}C\protect\textsubscript{32}.
b) $F_{\text{ah}}$ for Zr\protect\textsubscript{32}C\protect\textsubscript{31}.
In each figure, error bars represent the deviation of $F_{\text{ah}}$
from a DFT TI method reference (TU-TILD\cite{Duff2015}). c)
Vacancy anharmonicity, specified as an excess Gibbs free energy at
ambient pressure.
 \label{fig: pltFah}}}
\end{figure}
\par\end{center}

Our MEAM TI approach predicts $F_{\text{ah}}(V,T)$ to within
a target accuracy of 1 meV/at compared to DFT endpoint TI. This is
demonstrated in Fig. \ref{fig: pltFah}a. Perfect Zr\textsubscript{32}C\textsubscript{32}
is shown with error bars (TU-TILD reference) for 25 volumes and temperatures
up to the melting point. $F_{\text{ah}}(V,T)$ energies
are converged to sufficient precision that the error bars are in effect
systematic potential error bars. The mean absolute error (MAE) is
0.5 meV/at, with a mean signed deviation of 0.05 meV/at. The MAE values
at the lattice parameters $\{4.685,\,4.730,\,4.759,\,4.801,\,4.850\}$
\AA \ are $\{0.72,\,0.38,\,0.35,\,0.46,\,0.64\}$ meV/at. 
Errors resolved
at the temperatures $\{760,\,1900,\,2500,\,3200,\,3805\}$ K have
the MAE values $\{0.39,\,0.48,\,0.43,\,0.59,\,0.66\}$ meV/at. 

On the basis of adequately small errors for bulk ZrC, we propose using
the MEAM thermodynamic integration approach for more complex systems.
In this regard Zr\textsubscript{32}C\textsubscript{31} is a useful
test case for two reasons. The carbon-vacancy introduces complexity
in terms of physical interactions. It removes inversion symmetry at
 sites around the vacancy, so there are terms in the energy of odd-order in atomic displacements, previously
excluded by symmetry in perfect ZrC. Secondly, making free energy
predictions per vacancy increases computational complexity considerably due to the nature of statistical
error scaling for TI predictions on a per-vacancy basis. 

For the vacancy system Zr\textsubscript{32}C\textsubscript{31},
$F_{\text{ah}}(V,T)$ is shown in Fig.~\ref{fig: pltFah}b.
Obtaining comparable DFT TI values for systems with vacancies
like Zr\textsubscript{32}C\textsubscript{31} is prohibitively expensive
in general but we have computed a  DFT benchmark for Zr\textsubscript{32}C\textsubscript{31}
at $a=4.801$ \AA \ and $T=3200$ K. The MEAM thermodynamic integration
is found to overestimate the DFT TI reference $F_{\text{ah}}$ value
by $0.4$ meV/at, which is comparable to the MAE in the perfect bulk
ZrC.

\subsection{Computational cost\label{subsec: speedUp}}

In Table \ref{tab: speedUP-MTILD-TUTILD} timings are reported for
the MEAM-based TI in this work and TU-TILD (DFT) calculations. Both
methods compute the $F_{\text{ah}}(V,\,T)$ surface across
25 mesh points for a Zr\textsubscript{32}C\textsubscript{32} test
case. The MEAM approach does not have DFT TI and DFT TI up-sampling steps,
which account for the majority of the TU-TILD $F_{\text{ah}}$ cost. In the
MEAM approach, the main CPU-time overhead is high-quality DFT calculations on selected MD configurations, which are used create data to fit the MEAM potentials. 
Furthermore the
time required to optimize\cite{Duff2016} the MEAM potential with a large fitting set is substantially longer (ca. $\times10$), compared to a MEAM potential used in the intermediate TI steps in TU-TILD. Despite this the former scheme still gains
a factor of at least $\times10$ in efficiency overall due to having no DFT TI or DFT TI up-sample.
For anharmonic predictions where 1 meV/atom convergence is sufficient,
the TI method described in this work is likely to be a cost effective
choice for metals and alloys. It would be of interest to compare the efficiency of less specialised machine learned potentials\citep{Cheng2019,Behler2017,Grabowski} to the MEAM type as applied here, in terms of parameter fitting time, required training DFT data and potential compute time.

\begin{table}
\caption{\foreignlanguage{american}{Computer resources consumption for Zr\protect\textsubscript{32}C\protect\textsubscript{32}
test case calculation of $F_{\text{ah}}(V,T)$ on a $5\times5$
$\{V_{i},\,T_{i}\}$ mesh. Timings listed in CPU core-hours
and quasiharmonic-free-energy-job units ($1/t(F_{\text{qh}})$) using
the reference value $t(F_{\text{qh}})=4800$ core-hours. \label{tab: speedUP-MTILD-TUTILD}}}

\centering{}%
\begin{tabular}{cccccc}
\hline 
\noalign{\vskip\doublerulesep}
\multirow{2}{*}{Contributions } & \multicolumn{2}{c}{$t(F_{\text{ah}})$ (core-hrs)} & \selectlanguage{american}%
\selectlanguage{american}%
 & \multicolumn{2}{c}{$t(F_{\text{ah}})/t(F_{\text{qh}})$}\tabularnewline[\doublerulesep]
\cline{2-3} \cline{5-6} 
\noalign{\vskip\doublerulesep}
 & this work & TU-TILD & \selectlanguage{american}%
\selectlanguage{american}%
 & this work & TU-TILD\tabularnewline[\doublerulesep]
\hline 
\hline 
Fit set DFT MD  & $10^{4}$  & $10^{3}$ & \selectlanguage{american}%
\selectlanguage{american}%
 & 10 & 0.6\tabularnewline
MEAM fitting & $10^{3}$  & $10^{2}$ & \selectlanguage{american}%
\selectlanguage{american}%
 & 0.4 & 0.03\tabularnewline
MEAM TI & $10^{3}$  & $10^{2}$ & \selectlanguage{american}%
\selectlanguage{american}%
 & 0.4 & 0.1\tabularnewline
DFT TI & - & $10^{5}$ & \selectlanguage{american}%
\selectlanguage{american}%
 & - & 110\tabularnewline
DFT up-sampling & - & $10^{4}$ & \selectlanguage{american}%
\selectlanguage{american}%
 & - & 6\tabularnewline
\hline 
Total & $10^{5}$ & $10^{6}$ & \selectlanguage{american}%
\selectlanguage{american}%
 & 11 & 117\tabularnewline
\hline 
\end{tabular}
\end{table}

\section{Application of TI method to ZrC\protect $_x$ \label{sec: Application} }

\subsection{The character of anharmonicity in ZrC\protect$_x$\label{subsec: vacancyAnharmonicity}}

Prior to discussing the substoichiometric crystal, consider the anharmonic
contribution to the Helmholtz free energy of perfect ZrC, shown in
Fig. \ref{fig: pltFah}a. $F_{\text{ah}}(V,\,T)$ in
Zr\textsubscript{32}C\textsubscript{32} tends to be positive and
increase with temperature. This is because the anharmonic phase space
has a smaller volume within a given potential energy surface, giving
a positive anharmonic free energy term. A positive anharmonic contribution is similarly observed in other
extended systems,\cite{Glensk2014,Glensk2015,Zhang2018b} and is expected to be dependent on the presence of inversion symmetry.

$F_{\text{ah}}(V,\,T)$ in Zr\textsubscript{32}C\textsubscript{32}
increases with volume expansion. This contrasts with the result of harmonic
force constants, which typically become softer under tensile strain, increasing the entropy and \emph{decreasing} the free energy.
In terms of effective frequencies in perfect ZrC, quasiharmonicity
reduces frequencies with volume expansion whereas anharmonicity
in ZrC increases frequencies.

The anharmonic free energy of the Zr\textsubscript{32}C\textsubscript{31}
crystal is given in Fig. \ref{fig: pltFah}b. 
$F_{\text{ah}}(V,\,T)$ for Zr\textsubscript{32}C\textsubscript{31}
naturally appears similar to Zr\textsubscript{32}C\textsubscript{32}, since most atoms in Zr\textsubscript{32}C\textsubscript{31} are fully coordinated,
but the anharmonic free energy is less positive, for example, $F_{\text{ah}}(V,T)$ is lower by approximately $4$ meV/at at $a=4.759$ \AA \ and $T=3800$ K. 

To directly identify vacant-site anharmonicity we compute
\begin{align}
\label{eqn:excessF}
F_{\text{ah}}^{\text{excess}}=F_{\text{ah}}\left(\text{Zr}_{32}\text{C}_{31}\right)-F_{\text{ah}}\left(\text{Zr}_{32}\text{C}_{32}\right)\,,
\end{align}
which isolates the vacancy anharmonic contribution
by cancelling common contributions in Zr\textsubscript{32}C\textsubscript{32}
and Zr\textsubscript{32}C\textsubscript{31}. The anharmonicity of
a single vacant site $F_{\text{ah}}^{\text{excess}}$ is stronger
and qualitatively different in character to the anharmonicity per
site in $F_{\text{ah}}$ for Zr\textsubscript{32}C\textsubscript{32}.
$F_{\text{ah}}^{\text{excess}}$ for example typically exceeds $F_{\text{ah}}$ by 
more than an order of magnitude (approximately $\times20$),
and $F_{\text{ah}}^{\text{excess}}$ is negative whereas $F_{\text{ah}}$
is almost always positive.

As we are typically interested in ambient pressure thermodynamics, we can consider the nature of the excess anharmonic Gibbs free energy $G_{\text{ah}}^{\text{excess}}$ rather than Helmholtz  $F_{\text{ah}}^{\text{excess}}$. $G_{\text{ah}}^{\text{excess}}(T)$ in Fig. \ref{fig: pltFah}c illustrates the strength and sign of vacancy anharmonicity at ambient pressure. The large negative values of $G_{\text{ah}}^{\text{excess}}$ at high temperature can be simply rationalized. At high temperature the change in
thermal excursions, when atoms are near a vacancy, is larger than
predicted by harmonic springs, so the entropy is greater and free
energy less. In terms of the change in the anharmonic potential, the magnitude and
sign of $G_{\text{ah}}^{\text{excess}}$ are attributed to terms that start from
third-order in the potential Taylor expansion, rather than fourth-order
as in the perfect crystal with inversion symmetry.

\subsection{Vacancy volume, formation energy, and concentration\label{subsec: anharmonicVacancyEnergy}}

\begin{figure}[h]
\begin{centering}
\includegraphics[scale=0.56]{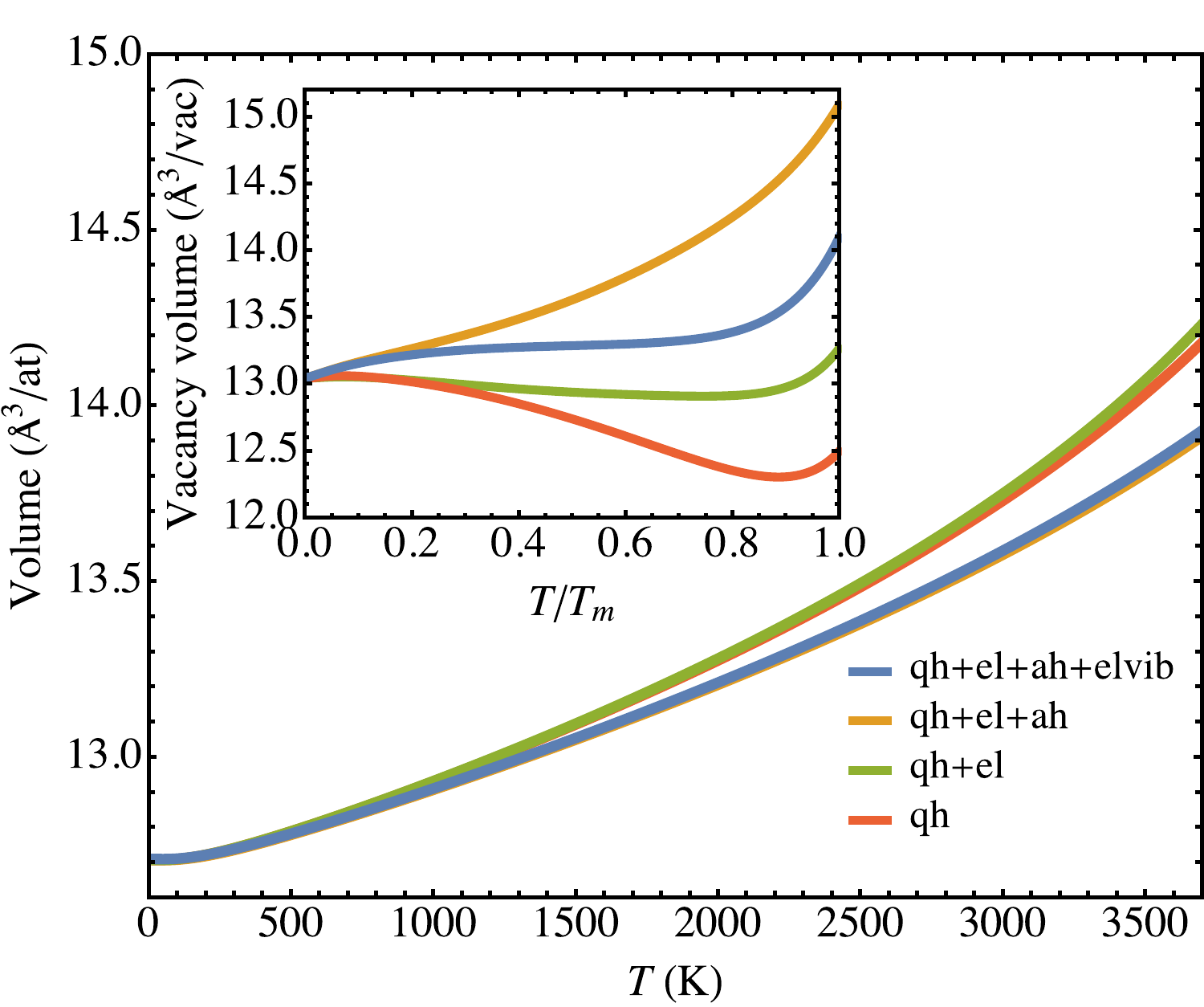}
\par\end{centering}
\caption{\foreignlanguage{american}{Thermal expansion of ZrC, $V(T)$, and inset, the vacancy formation volume $v^f(T)$. \label{fig: pltVolExpansion}}}
\end{figure}

\begin{figure}[h]
\begin{centering}
\includegraphics[scale=0.55]{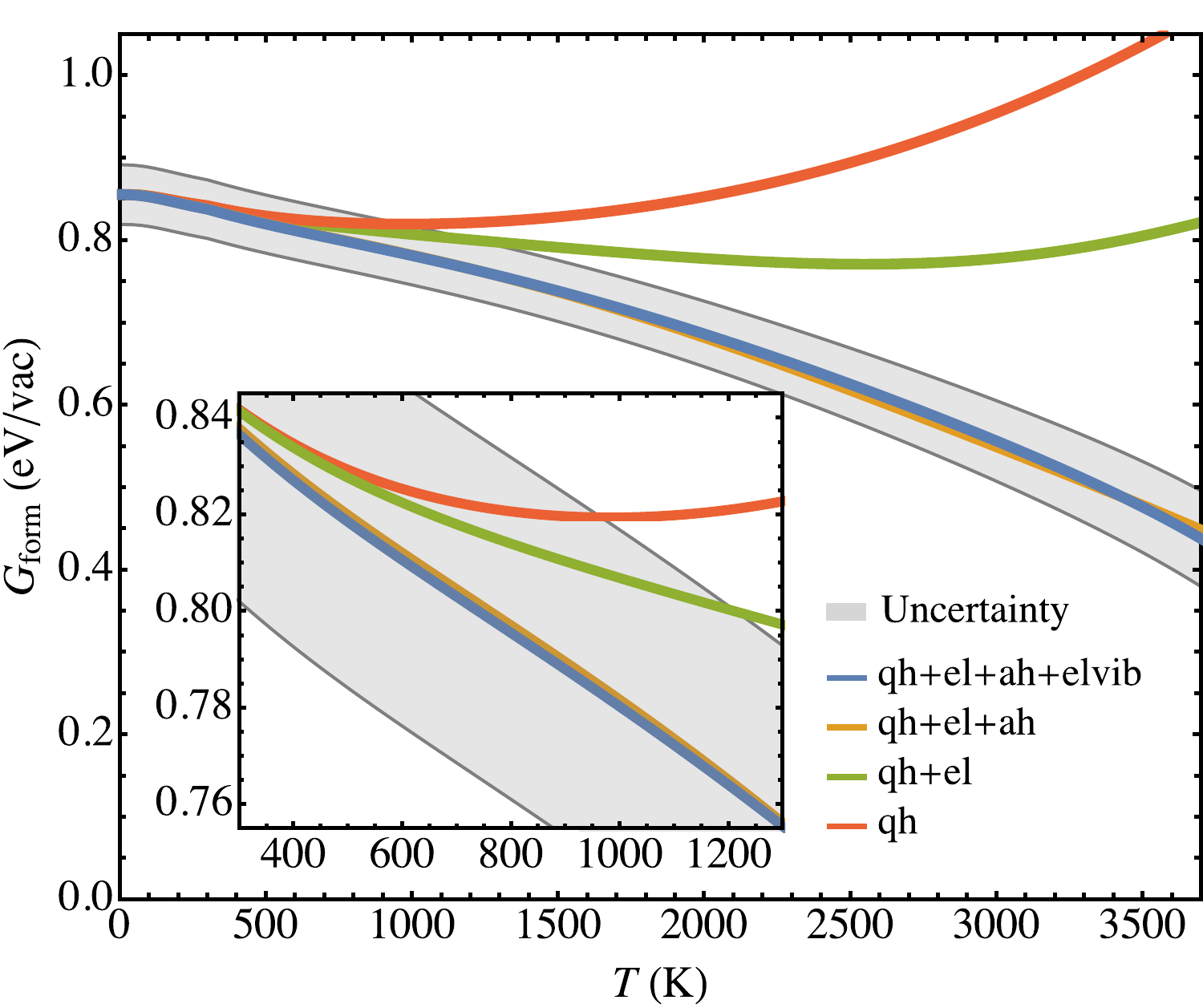}
\par\end{centering}
\caption{\foreignlanguage{american}{Gibbs free energy of carbon vacancy formation in ZrC versus temperature.
\emph{Inset}: The onset temperature of non-negligible anharmonicity.  The error bar shown is for the TI method used to determine $F_\text{ah}$, from MEAM-DFT potential errors (assuming no cancellation between Zr\textsubscript{32}C\textsubscript{32} and  Zr\textsubscript{32}C\textsubscript{31}), and statistical convergence..
\label{fig: gFormWithInset}}}
\end{figure}

\begin{figure}[h]
\begin{centering}
\includegraphics[scale=0.55]{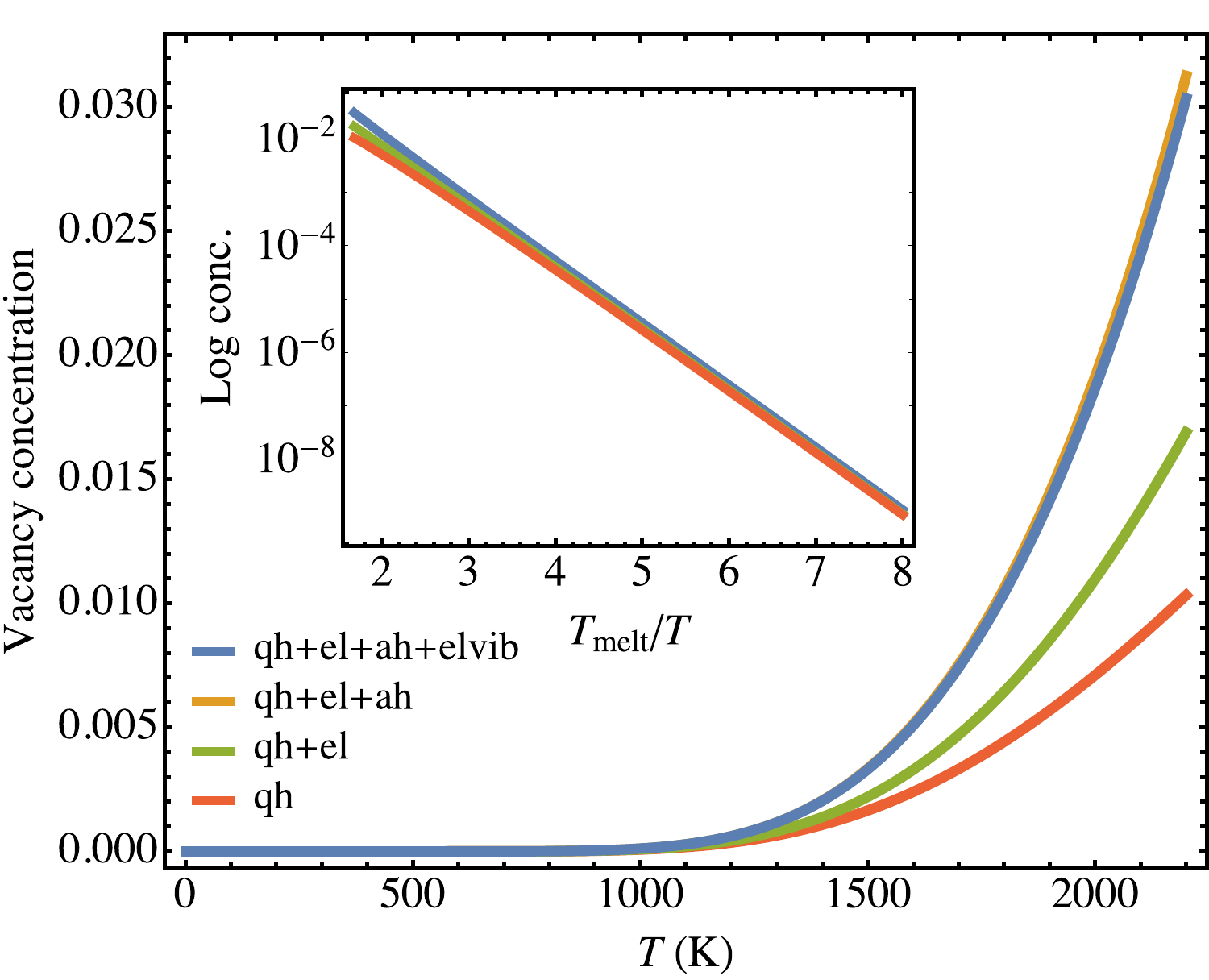}
\par\end{centering}
\caption{\foreignlanguage{american}{Carbon vacancy concentration in ZrC\protect$_x$ versus
temperature, up to a concentration of $c_{\text{vac}}=1/32$. \emph{Inset}:
Log vacancy concentration versus $T_{m}/T$. \label{fig: vacConcWithInset}}}
\end{figure}

\begin{table}
\caption{\foreignlanguage{american}{Concentration of vacancies ($c_{\text{vac}}$) in ZrC\protect$_x$
in carbon atom \%. \label{tab: temperature-vacancyConc}}}

\centering{}%
\begin{tabular}{ccc}
\hline 
\noalign{\vskip\doublerulesep}
\multirow{2}{*}{$T$ (K) } & \multicolumn{2}{c}{$c_{\text{vac}}$ (C at. \%)}\tabularnewline[\doublerulesep]
\noalign{\vskip\doublerulesep}
 & this work & CALPHAD\cite{FernandezGuillermet1995}\tabularnewline[\doublerulesep]
\hline 
300 & $9\times10^{-13}$ & $1\times10^{-15}$\tabularnewline
500 & $6\times10^{-7}$ & $1\times10^{-9}$\tabularnewline
1000 & 0.01 & 0.003\tabularnewline
1500 & 0.33 & 0.06\tabularnewline
2000 & 1.87 & 0.33\tabularnewline
2200  & 3.03 & 0.51\tabularnewline
\hline 
\end{tabular}
\end{table}

The thermal expansion of the ZrC atomic volume ($V$) is shown in Fig.\,\ref{fig: pltVolExpansion}, alongside the vacancy formation volume, $v^f(T)=\Omega-NV$, where $\Omega$ is the volume of an  $N$ atom defective ZrC$_x$ cell. For ZrC, predictions at the qh+el+ah level of theory reproduce the thermal expansion reported in an earlier theoretical work,\citep{Duff2015} while including electron-vibration coupling provides a small-to-negligible additional enhancement, evident in Fig.\,\ref{fig: pltVolExpansion}. 

The volume of a vacant carbon site at $T=298$ K is $v^f=13.1$ \AA$^3$/vac, which is +3.2 \% or +0.40 \AA$^3$/site larger than the corresponding atomic volume $V$ for the perfect crystal. This means the lattice of a ZrC$_x$ crystal initially expands for $x<1$, with the lattice parameter $a$ increasing by +0.001 \AA$\,$  from $x=1$ to $x=0.97$ in our 298-K calculations. This is at odds with recent measurements by Nakayama et al.\citep{Nakayama2017} who report a monotonic, apparently linear trend, but is supported by other experimental work in which the ZrC$_x$ lattice parameter is a concave function of carbon substoichiometry.\citep{Sara1900, Schonfeld2017, Katoh2013, Jackson2011} For example according to Sara,\citep{Sara1900} the maximum volume occurs at composition ZrC$_{0.90}$ with an $a$ value some +0.004 \AA $\,$ greater than in ZrC$_{0.98}$. To first order, the gradient is ca. +0.0011 Ang/C at. \%, compared to +0.0005 Ang/C at. \% in our work. It is important to stress that this is quite a subtle effect, and that it is temperature dependent. For $T>2200$ K our computed lattice constant \emph{decreases} from $x=1$ to $x=0.97$. 

As a final comment on thermal expansion, we note the temperature dependence of $v^f$ is somewhat complicated (Fig.\,\ref{fig: pltVolExpansion} inset). Quasiharmonic volume-dependent frequencies and electron-vibration coupling make the vacancy volume smaller generally, whereas anharmonic and electron thermal excitations increase it. In each instance, at high temperature such as $T\ge0.75\,T_m$, these effects are comparable in size to the 0-K outward relaxation of the Zr atoms around the vacancy; ZrC bonds normal to the vacancy surface are squeezed by -0.08 \AA$\,$ compared to bonds of length $d\text{(Zr-C)}=2.328$ \AA$\,$ in the perfect crystal.

The energy to form a carbon vacancy in ZrC is considered in terms of a Gibbs free energy computed as follows:
\begin{equation}
G_{\text{form}}^{\text{}}=G\left(\text{Zr}_{32}\text{C}_{31}\right)+\mu(\text{C})-G\left(\text{Zr}_{32}\text{C}_{32}\right)\,.
\label{eq: gForm}
\end{equation}
In this expression $\mu(\text{C})$ is the chemical potential to remove an atom of carbon from ZrC and place it in a carbon reservoir. The reference state of carbon is taken to be that of graphite, which for 0 K to 298 K, is computed by quasiharmonic DFT for diamond, with a 0-K experimental correction to graphite. At higher temperatures, the experimental parameterization of the graphite free energy is used, in the form of the Gustafson\cite{Gustafson1986} assessment. This provides a diamond chemical potential that includes all contributions (e.g. anharmonicity) and is consistent with the DFT calculated free energies, while avoiding expensive calculations for graphite. Further details and an expression for $\mu(\text{C})$ are given in the Appendix.

The error bar in $G_{\text{form}}^{\text{}}(T_m)$, due to TI statistical precision and MEAM systematic potential error, is 60 meV/vacancy. This value assumes no cancellation in the systematic potential error between Zr\textsubscript{32}C\textsubscript{32} and Zr\textsubscript{32}C\textsubscript{31}, and is therefore an upper limit. As most sites in Zr\textsubscript{32}C\textsubscript{31} are fully-coordinated and bulk-like, and the same MEAM potential is used to describe perfect Zr\textsubscript{32}C\textsubscript{32} and Zr\textsubscript{32}C\textsubscript{31}, partial cancellation of the systematic potential error is expected. In the limit of anharmonicity being a site-localised property, the systematic potential error would only arise from   the six under-coordinated nearest-neighbours to the vacancy in Zr\textsubscript{32}C\textsubscript{31}, and the corresponding non-matching seven sites in Zr\textsubscript{32}C\textsubscript{32}. In this case the total error is less than 10 meV/vacancy.

$G_{\text{form}}^{\text{}}(T)$ is shown in Fig.\,\ref{fig: gFormWithInset},
including quasiharmonic, electronic, electron-vibration, and anharmonic contributions. Above ca. $1000\,$K ($T/T_{m}\approx0.3$) the anharmonic contribution
can no longer be regarded as negligible, and above ca. $2000\,$K ($T/T_{m}\approx0.5$)
accounting for anharmonicity is critical to qualitatively describe
the ZrC vacancy formation energy. With respect to a quasiharmonic reference,
Fig. \ref{fig: gFormWithInset} shows that electronic entropy lowers
the formation energy, and that anharmonicity substantially lowers the Gibbs formation energy further, while the electron-vibration contribution is much smaller.
In the final predictions, which include the quasiharmonic, electronic, electron-vibration, and anharmonic effects, $G_{\text{form}}^{\text{}}$ is almost linear in temperature, and decreases by approximately 10 meV with every increase in temperature by 1000 K. This rate of decrease is similar to reports in other materials such as aluminum and nickel.\cite{Glensk2014,Gong2018}

The vacancy concentration in contact with graphite is computed with an ideal solution model 
\begin{equation}
\frac{c_{\text{vac}}}{1-c_{\text{vac}}}=\text{exp}\left(-\frac{G_{\text{form}}}{k_\text{B} T}\right)\,,
\label{eq: cVac}
\end{equation}
and is shown in Fig. \ref{fig: vacConcWithInset}. Anharmonicity favors
vacancy formation by making $G_{\text{form}}^{\text{}}$ smaller, increasing
$c_{\text{vac}}$ by a factor of two compared to predictions
at the $F_{\text{qh}}+F_{\text{el}}$ level. The effect of electron-vibration coupling on vacancy concentration is marginal in this material. Specific values of $c_{\text{vac}}(T)$
are shown in Table \ref{tab: temperature-vacancyConc}, up to a temperature of $T=2200$
K, which is when the predicted concentration reaches our operative dilute
limit of one vacancy per supercell ($c_{\text{vac}}=1/32$ for ZrC$_x$).

In Table \ref{tab: temperature-vacancyConc} the $c_{\text{vac}}$
values from the CALPHAD assessment are consistently
lower than our $c_{\text{vac}}$ values. \cite{FernandezGuillermet1995}
Despite the power of the CALPHAD method for ZrC$_x$,\cite{FernandezGuillermet1995}
uncertainties can arise from insufficient experimental data, and the
limitations that exist due to the non-physical interaction terms the
methodology assumes. 
At $T=2000$ K the CALPHAD value is $c_{\text{vac}}=0.3$ C at. \%,\cite{FernandezGuillermet1995}
compared to $c_{\text{vac}}=1.9$ C at. \% in this work.

Our predictions have quantum mechanical many-body errors from the
LDA exchange-correlation treatment we use to describe ZrC. While the GGA
has been shown to be less suitable to describe ZrC at high temperature
than LDA,\cite{Duff2015}  it is instructive to consider the vacancy
formation energy from both exchange-correlation treatments, in order to gauge sensitivity. At $T=0$ K the GGA vacancy formation energy is less than the LDA value by some
0.2 eV/vacancy (without zero-point corrections and dilute limit supercell
extrapolation), indicating a GGA predicted concentration is greater.
Quantitative predictions of the non-local quantum many-body error
at high temperature is beyond the scope of this work, but experience suggests that the LDA and GGA (PBE) functionals bracket the exact result.\citep{Grabowski2015}

In this work we confine our predictions to temperatures at which concentrations
do not exceed one vacancy per supercell. This should minimize lattice
many-body errors, however we note that vacancy-vacancy interactions
are expected to be mainly repulsive,\cite{Zhang2015} and that other
entities on the carbon sub-lattice such as Frenkel defects will decrease the vacancy configuration
space.\cite{Mellan2018} These effects are expected to moderate $c_{\text{vac}}$, to values lower than ideal,
to an extent that increases with temperature. 

\begin{figure}[h]
\begin{centering}
\includegraphics[scale=0.55]{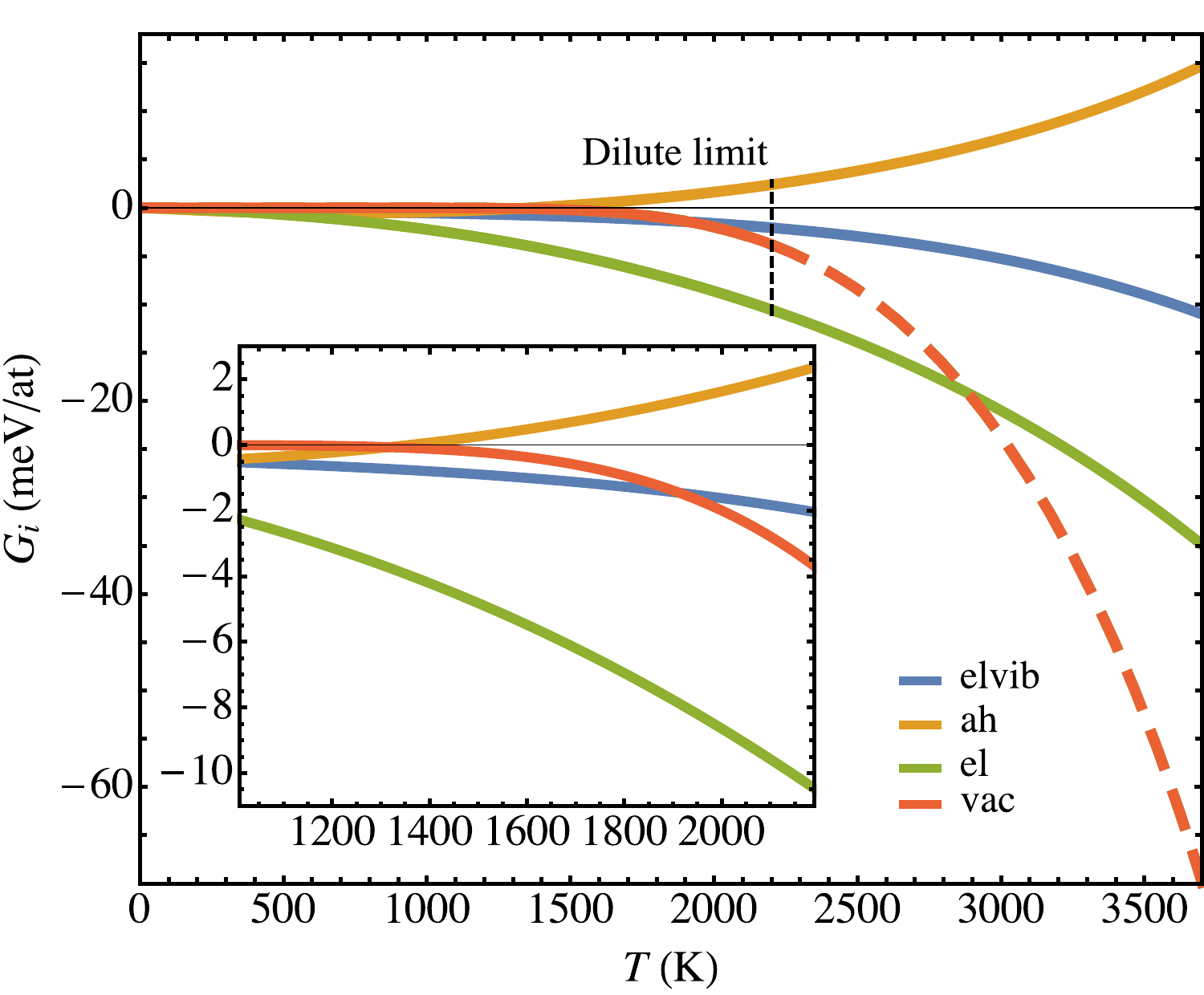}
\par\end{centering}
\caption{\foreignlanguage{american}{ZrC\protect$_x$ Gibbs free energy contributions with
respect to a quasiharmonic reference. \emph{Inset}: Gibbs free energy
shown from 1200 K to the dilute vacancy concentration ($1/32$) at
2200 K. \label{fig: freeEnergyContributionsRelQhaInset}}}
\end{figure}

\begin{figure}[h]
\begin{centering}
\includegraphics[scale=0.55]{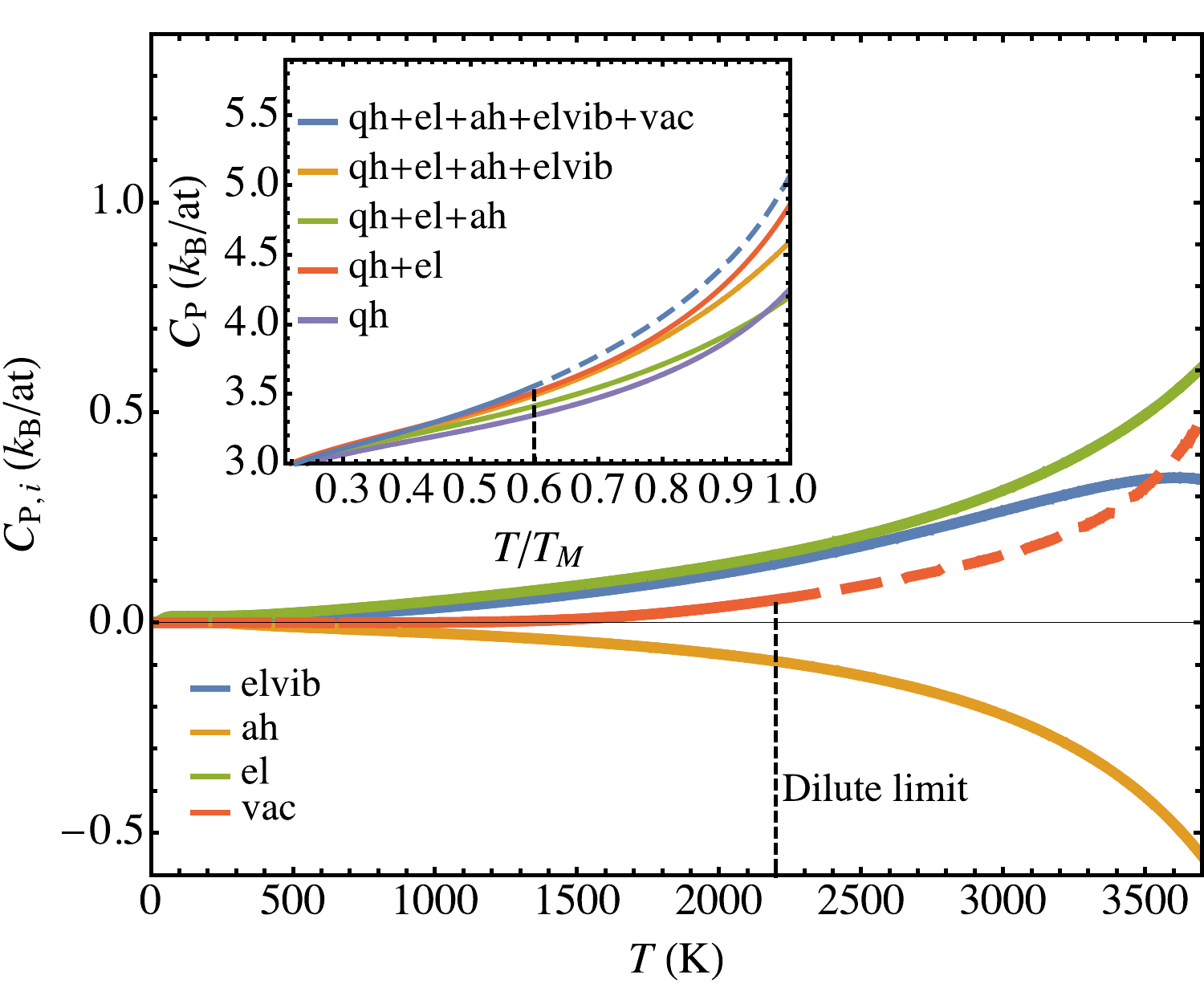}
\par\end{centering}
\caption{\foreignlanguage{american}{ZrC\protect$_x$ constant-pressure heat capacity relative
to a quasiharmonic reference. \emph{Inset}: heat capacity at different
levels of theory, shown from the Debye temperature ($C_{\text{p}}(T_{\text{Debye}})=3$
\emph{k}\protect\textsubscript{B}) to the melting point ($T_{m}=3700$
K). \label{fig: pltCpRelative}}}
\end{figure}

\subsection{Free energy and heat capacity of ZrC\protect$_x$\label{subsec: testApplicationZrCx-heatCapacity}}

The basic excitation mechanisms that determine the thermal properties of ZrC$_x$  are discussed relative to a quasiharmonic reference system.  Formulae for the referenced Gibbs free energies at ambient pressure, $G_{\text{el}}$, $G_{\text{el-vib}}$, $G_{\text{ah}}$ and 
$G_{\text{vac}}$, are listed in the Appendix.
Each is shown as a function of temperature in Fig. \ref{fig: freeEnergyContributionsRelQhaInset}.

At high temperature the magnitude of the electron-vibration contribution to the Gibbs free energy is less than the anharmonic contribution, which is in turn less than the electronic contribution. Partial cancellation occurs as $G_{\text{el}}$ and $G_{\text{el-vib}}$ are negative whereas $G_{\text{ah}}$ is positive in this material. The vacancy contribution $G_{\text{vac}}$ is the smallest of the four contributions up to 1900 K, but beyond the dilute vacancy concentration temperature of $T=2200$ K, $G_{\text{vac}}$ increases considerably. Extrapolating to higher temperatures, the vacancy contribution appears to become the largest of all above $3000$ K. Note however that above $T=2200$ K the value of $G_{\text{vac}}$ is presented as indicative only, and is represented in Fig. \ref{fig: freeEnergyContributionsRelQhaInset} with a dashed line, as it exceeds the thermodynamic limitations of our dilute solution model.

The different  ZrC$_x$ heat capacity contributions, relative to the quasiharmonic system, are shown in Fig. \ref{fig: pltCpRelative}. The anharmonic term $C_{\text{P},\,\text{ah}}$ is negative and the electronic one $C_{\text{P},\,\text{el}}$ is positive, with each similar in absolute value near $T_m$. It is somewhat interesting to consider the extent to which $C_{\text{P},\,\text{ah}}$ and $C_{\text{P},\,\text{el}}$ cancellation is coincidental in ZrC$_x$ or a manifestation of a generic feature. In ordinary metals $C_{\text{P},\,\text{el}}$ is \emph{a priori} positive, and for ordinary high-symmetry crystals, $C_{\text{P},\,\text{ah}}$ is negative at high temperature.\cite{Glensk2015} Some cancellation of $C_{\text{P},\,\text{ah}}$ and $C_{\text{P},\,\text{el}}$ is therefore regarded to be likely in conducting systems such as the refractory ceramic ZrC. However, as the extent of cancellation depends on the magnitude of each contribution, for which we are unaware of a direct physical relation, we conclude cancellation is mostly coincidental.

In ZrC$_x$, $C_{\text{P},\,\text{ah}}$, $C_{\text{P},\,\text{el}}$, and $C_{\text{P},\,\text{el-vib}}$
are all individually larger than the vacancy contribution. For example at 2200 K, $ C_{\text{P,\,vac}}=+0.06$ \emph{k}\textsubscript{B}/atom,
compared to $ C_{\text{P,\,el}}=+0.16$ \emph{k}\textsubscript{B}/atom, $C_{\text{P,\,el-vib}}=+0.14$ \emph{k}\textsubscript{B}/atom 
and $C_{\text{P,\,ah}}=-0.09$ \emph{k}\textsubscript{B}/atom. Extending the vacancy model  beyond the dilute limit with the dashed line in Fig. \ref{fig: pltCpRelative} indicates $C_{\text{P,\,vac}}$ is comparable to the positive contributions of $C_{\text{P},\,\text{el}}$, and $C_{\text{P},\,\text{el-vib}}$ near $T_m$.  Consequently we suggest the physical origin of the steep increase in heat capacity in ZrC$_x$\cite{Savvatimskiy2017} is a combination of electron thermal excitations, electron-vibration coupling, and structural excitations on the carbon sub-lattice, predominantly the constitutional carbon vacancies that have been the focus of this paper, although there are also stoichiometry conserving intrinsic carbon Frenkel defects, which are beyond the scope of this work but discussed elsewhere.\cite{Mellan2018,Savvatimskiy2017} Finally, it is interesting to note that while anharmonicity is the only term that suppresses $C_\text{P}$ in Fig. \ref{fig: pltCpRelative}, the enhancing  effects of the vacancy contribution are mainly due to the indirect effect of anharmonicity, insomuch as anharmonicity by lowering $G_\text{form}$ enhances the population of vacancies (e.g. by a factor of $\times2$ at 2200 K  as shown in Fig. \ref{fig: vacConcWithInset}).

\section{Conclusions}

\subsubsection*{Thermodynamic integration approach}

We have described a thermodynamic integration method to calculate
the anharmonic free energy of a crystal to DFT accuracy. In our benchmark system of ZrC the approach achieves average target
precision better than 1 meV/atom and 60 meV/vacancy, comparable to more expensive
DFT-based thermodynamic integration schemes. The method transfers
the burden of computation from converging random statistical errors
to minimizing systematic potential errors. With sufficient care to
minimize potential errors, precise calculations can be made to compute
quantities such as the anharmonic vacancy formation energy, with modest
computing resources.

\subsubsection*{Application to ZrC\protect$_x$}

Anharmonicity increases the concentration of vacancies in ZrC$_x$.
At 2000 K vacancies increase from $1.1$ to $1.9$\,C\,at.\,\% due to explicit
anharmonicity. 
The ZrC$_x$ heat capacity is enhanced by electron thermal excitations, electron-vibration coupling and vacancies on the carbon sub-lattice, and suppressed by anharmonicity. For example, $C_{\text{P,\,el}}(T=2200\,\text{K})=+0.16$ \emph{k}\textsubscript{B}/atom, $C_{\text{P,\,el-vib}}(T=2200\,\text{K})=+0.14$ \emph{k}\textsubscript{B}/atom, $C_{\text{P,\,vac}}(T=2200\,\text{K})=+0.06$ \emph{k}\textsubscript{B}/atom, and $C_{\text{P,\,ah}}(T=2200\text{K})=-0.09$ \emph{k}\textsubscript{B}/atom.  The sharp increase in the heat capacity at high temperature is attributed to electronic and electron-vibration effects along with the thermal excitation of structural defects.

\section{Acknowledgements}

T.A.M. and M.W.F. acknowledge computational support from the UK national high
performance computing service, ARCHER, for which access was obtained
via the UKCP consortium and funded by EPSRC grant EP/P022561/1, and 
for computational support from the UK Materials and
Molecular Modelling Hub, which is partially funded by EPSRC grant EP/P020194. 
T.A.M. and M.W.F. acknowledge the
financial support of EPSRC Programme Grant No. EP/K008749/1 Material
Systems for Extreme Environments (XMat), and Grant No. EP/M018563/1 Carbides
for Future Fission Environments (CAFFE). M.W.F. is grateful for support
from the Alexander von Humboldt-Stiftung award. A.I.D. acknowledges
support from the STFC Hartree Centre\textquoteright s Innovation:
Return on Research programme, funded by the UK Department for Business,
Energy \& Industrial Strategy. 
B.G. acknowledges funding from the European Research Council (ERC) under the European Union Horizon 2020 research and innovation programme (grant agreement No. 639211). 

\cleardoublepage{}

\section{References}

\selectlanguage{american}%
\bibliographystyle{apsrev4-1}

\bibliography{ZrC_TI_manuscript.bib}

\appendix

\section*{Appendix\label{sec:Appendix}}

\section*{Gibbs free energy contributions\label{sec: gibbsFreeEnergies}}
To quantify different thermal excitations beyond the quasiharmonic level of theory, quasiharmonic-referenced Gibbs free energies are reported as
\[
G_{\text{ah}}=\underset{V}{\text{min}}\left[F_{\text{ah}}+F_{\text{qh}}+E_{0}+pV\right]-\underset{V}{\text{min}}\left[F_{\text{qh}}+E_{0}+pV\right]\,,
\]
\[
G_{\text{el-vib}}=\underset{V}{\text{min}}\left[F_{\text{el-vib}}+F_{\text{qh}}+E_{0}+pV\right]-\underset{V}{\text{min}}\left[F_{\text{qh}}+E_{0}+pV\right]\,,
\]
and 
\[
G_{\text{el}}=\underset{V}{\text{min}}\left[F_{\text{el}}+F_{\text{qh}}+E_{0}+pV\right]-\underset{V}{\text{min}}\left[F_{\text{qh}}+E_{0}+pV\right]\,.
\]
Similarly $G_{\text{vac}}$ is the Gibbs free energy associated with a concentration of vacancies ($c_\text{vac}$) in ZrC$_x$, again referenced to the quasiharmonic system. This is defined by writing the total Gibbs free energy of ZrC$_x$ at the full level of theory as $G=G^{\text{perf}}-c_{\text{vac}}k_\text{B}T$. In this expression $G^{\text{perf}}=\underset{V}{\text{min}}\left[F_{\text{ah}}+F_{\text{el-vib}}+F_{\text{qh}}+F_{\text{el}}+E_{0}+pV\right]$, and $c_\text{vac}$ is the equilibrium concentration of vacancies, which has been computed from the Arrhenius ideal solution model introduced in Eqn. (\ref{eq: cVac}). To compute $G_{\text{vac}}$, the vacancy part ($G-G^{\text{perf}}$) is referenced to the quasiharmonic system:
\begin{align*}
G_{\text{vac}} & =\left(G-G^{\text{perf}}\right)-\left(G_{\text{qh}}^{\,}-G_{\text{qh}}^{\text{perf}}\right)\\
 & =-\left(c_{\text{vac}}-c_{\text{vac}}^{\text{qh}}\right)k_\text{B}T\,,
\end{align*}
which is equivalent to the difference in equilibrium vacancy concentrations at the full and quasiharmonic levels of theory. Note for completeness, the quasiharmonic reference system terms are defined as follows: $G_{\text{qh}}=G_{\text{qh}}^{\text{perf}}-c_{\text{vac}}^{\text{qh}}k_\text{B}T$, with $G_{\text{qh}}^{\text{perf}}=\underset{V}{\text{min}}\left[F_{\text{qh}}+E_{0}+pV\right]$, and $c_{\text{vac}}^{\text{qh}}$ is the ideal solution model equilibrium concentration, with exponent $G_\text{form}$ (Eqn. (\ref{eq: gForm})) calculated at the quasiharmonic level.
\section*{Graphite chemical potential\label{sec:chemicalPotential}}
The ZrC vacancy formation energy has been calculated with respect
to a graphite chemical potential of the form
\begin{widetext}
\[
\mu(\text{C})=\begin{cases}

G_{\text{diamond}}(T)+\left(H_{\text{graphite}}-H_{\text{diamond}}\right) 

& 0\,\text{K}< T\le T_\text{stn}\,\\

\sum_{-3\le i\le2}a_{i}(T^i-T_\text{stn}^i)+a_{3}\,[T\,\text{ln}\,(T)-T_\text{stn}\,\text{ln}\,(T_\text{stn})]+ 
G_{\text{diamond}}(T_\text{stn})+\left(H_{\text{graphite}}-H_{\text{diamond}}\right) 

 & T_\text{stn}\,< T\le T_{m}\,.

\end{cases}\,
\]
The coefficients for $T>T_\text{stn}$, which is the CALPHAD standard state temperature $T_\text{stn}=298.15$ K, are set according to the Gustafson experimental
free energy parameterization:\cite{Gustafson1986}
\[
a_{[-3,\,3]}=\{1.2\times10^{10},\,-2.643\times10^8,\,2562600,\,-17369,\,170.73,\,-4.723\times10^{-4},\,-24.3\}\,.
\]
\end{widetext} At low temperatures ($T\le 298.15$ K) where the graphite parameterization
is unavailable, $\mu(\text{C})$ is continued using a DFT-calculated diamond potential, $G_{\text{diamond}}(T)$.
This quasiharmonic diamond potential at low-temperature is transformed to a graphite chemical potential by a correction equal to the $0$ K enthalpy difference $\left(H_{\text{graphite}}-H_{\text{diamond}}\right)=-0.03$
eV/atom.

\section*{Correlation time\label{sec: scatteringTimes}}
The correlation time $\tau$, which is used to estimate error scaling and determine statistical precision using stratified systematic sampling, is the integrated correlation time\cite{Janke2002}
\[
\tau(T,V,\lambda)\equiv \tau_{\Delta U, \text{int}}\,.
\]
$\tau_{\Delta U, \text{int}}$ is estimated by\cite{Janke2002}
\[
\tau_{\Delta U, \text{int}} = \left(\frac{1}{2} + \sum_{k=1}^N A(k)\right)\Delta t\,,
\]
with time-step  $\Delta  t = 1$ fs, and autocorrelation function, $A(k)$, given by
\[
A(k) = \frac{\langle \Delta U_i \Delta U_{i+k} \rangle - \langle \Delta U_i \rangle \langle \Delta U_{i} \rangle}{\langle \Delta U_i^2 \rangle - \langle \Delta U_i \rangle \langle \Delta U_{i} \rangle}\,.
\]
The calculated correlation time ranges from $9-13$ fs, depending weakly on the arguments of $\tau(T,V,\lambda)$.

\hfill

\section*{TI quadrature error\label{sec: errorTI}}
\begin{figure}[h]
\begin{centering}
\includegraphics[scale=0.55]{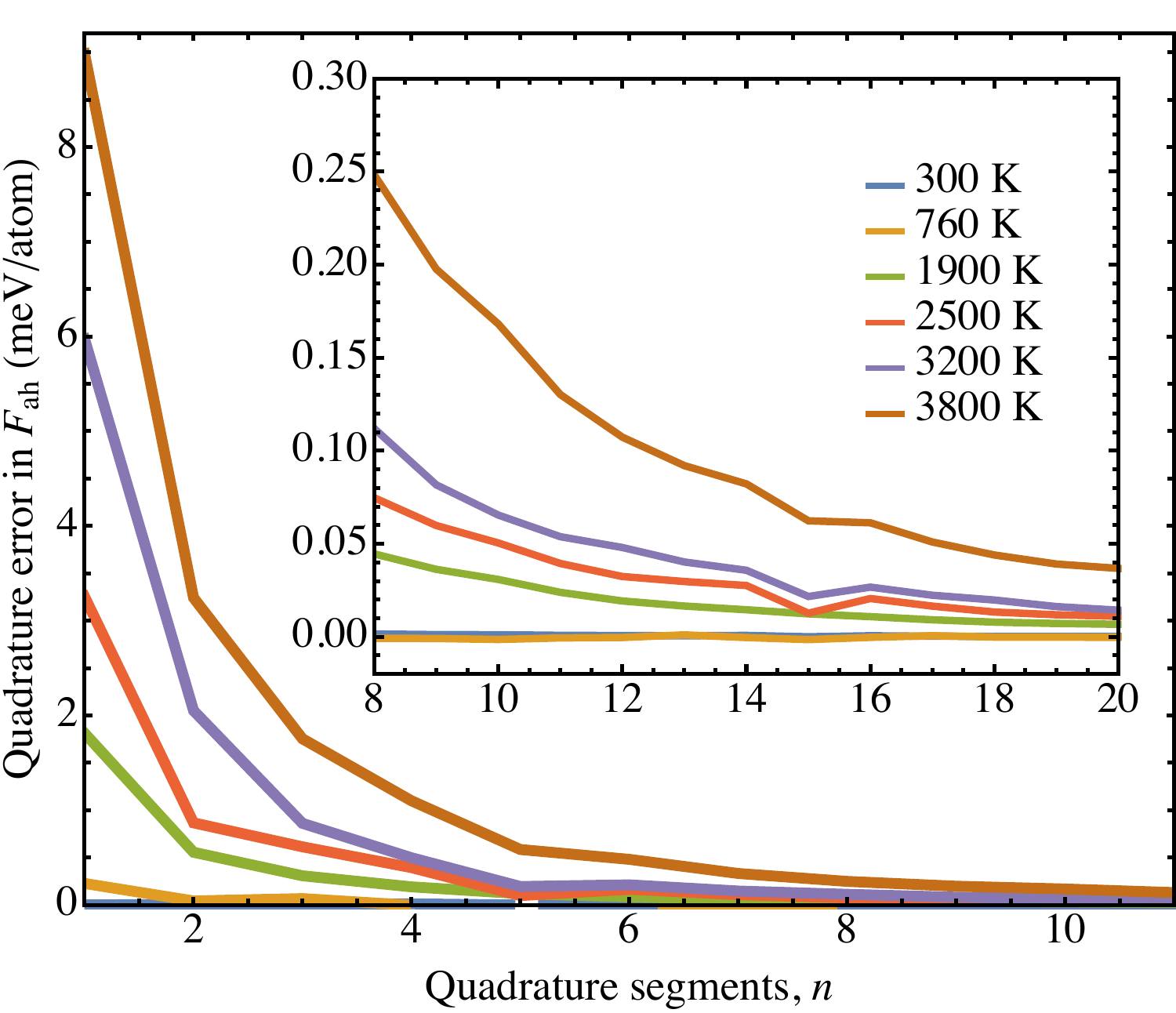}
\par\end{centering}
\caption{\foreignlanguage{american}{Thermodynamic integration error in $F_\text{ah}$ as function of the number of quadrature segments $n$ in $\lambda_i=i/n$.\label{fig: pltErrorTI}}}
\end{figure}

The quadrature error in $F_\text{ah}$ as a function of the number of integral sampling points ($\lambda_i$ values) is shown in Fig. \ref{fig: pltErrorTI}. At low temperatures when the system is nearly harmonic, the integrand $\left\langle \partial_{\lambda}E_{\text{mix}}(\lambda)\right\rangle _{\lambda}$ is small and almost independent of $\lambda$, and therefore a large number of $\lambda_i$ samples is unnecessary. At high temperature when $\left\langle \partial_{\lambda}E_{\text{mix}}(\lambda)\right\rangle _{\lambda}$ is curvier (see non-linearity in Fig. \ref{fig: pltdUdL}), sufficient sampling of the integrand is critical to obtain sub-meV/at numerical precision. $F_\text{ah}$ is determined in this work by sampling $\left\langle \partial_{\lambda}E_{\text{mix}}(\lambda)\right\rangle _{\lambda}$ at $10$ intervals or 11 points ($\lambda_i=i/10$). The associated error shown in Fig. \ref{fig: pltErrorTI} is less than $0.1$ meV/at for $T\le3200$ K, and ca. $0.2$ meV/at at $T_m$.

\end{document}